\documentstyle[11pt,aas2pp4,psfig]{article}

\def\etal{et al.~}

\def\hii{\ion{H}{2}}

\def\heii{\ion{He}{2}}

\def\hbeta{\ifmmode {\rm H{\beta}} \else $\rm H{\beta}$\fi}
\def\masl{\ifmmode  {\rm M_{\sun}yr^{-1}} \else ${\rm M_{\sun}yr^{-1}}$\fi}

\def\mdot{\ifmmode  \dot{M} \else $\dot{M}$\fi}
\def\msun{\ifmmode M_{\odot} \else $M_{\odot}$\fi}
\def\vinf{\ifmmode v_{\infty} \else $v_{\infty}$\fi}
\def\teff{\ifmmode T_{\rm eff} \else $T_{\rm eff}$\fi}
\def\logg{\ifmmode \log g \else $\log g$\fi}
\def\loggeff{\ifmmode \log g_{\rm eff} \else $\log g_{\rm eff}$\fi}
\def\rstar{\ifmmode R_{\star} \else $R_{\star}$\fi}
\def\lstar{\ifmmode L_{\star} \else $L_{\star}$\fi}
\def\mstar{\ifmmode M_{\star} \else $M_{\star}$\fi}
\def\rsun{\ifmmode R_{\odot} \else $R_{\odot}$\fi}
\def\lsun{\ifmmode L_{\odot} \else $L_{\odot}$\fi}
\def\12c16o{$^{12}{\rm C}\left(\alpha,\gamma\right)^{16}{\rm O}$}
\def\kms{\ifmmode {\rm km \;s^{-1}} \else $\rm km \;s^{-1}$\fi}
\def\nlte{non--LTE}

\begin{document}

\title{About the initial mass function and \heii\ emission in young starbursts}
\author{Daniel Schaerer\altaffilmark{1,2,3}}
\altaffiltext{1}{Space Telescope Science Institute, 3700 San Martin Drive, Baltimore,
	MD 21218}
\altaffiltext{2}{Geneva Observatory, CH-1290 Sauverny, Switzerland}
\altaffiltext{3}{Observatoire Midi-Pyr\'en\'ees, 14 Av. Edouard Belin,
	F-31400 Toulouse, France}
\authoremail{schaerer@stsci.edu}

\begin{abstract}
We demonstrate that it is crucial to account for the evolution 
of the starburst population in order to derive reliable numbers of 
O stars from integrated spectra for burst ages $t > 2 - 3$ Myr.
In these cases the method of Vacca \& Conti 
(1992) and Vacca 
(1994)
{\em systematically underestimates} the number of O stars.
Therefore the current WR/O number ratios in Wolf--Rayet (WR) galaxies 
are overestimated. This questions recent claims about flat 
IMF slopes ($\alpha \sim$ 1--2) in these objects.
If the evolution of the burst is properly treated we find that
the observations are indeed compatible with a Salpeter IMF,
in agreement with earlier studies.

Including recent predictions from \nlte, line blanketed 
model atmospheres which account for stellar winds, we synthesize the
nebular and WR \heii\ $\lambda$ 4686 emission in young
starbursts.
For metallicities $1/5 \, Z_\odot \le Z \le Z_\odot$ we predict
a {\em strong nebular} \heii\ {\em emission due to a significant fraction 
of WC stars} in early WR phases of the burst.
For other metallicities broad WR emission will always dominate 
the \heii\ emission. 
Our predictions of the nebular \heii\ 
intensity 
agree well with the observations in WR galaxies and 
an important fraction of the giant \hii\ regions where nebular 
\heii\ is detected.
We propose further observational tests of our result.
 
\end{abstract}


\twocolumn

\section{IMF determinations in young starbursts}
To derive the initial mass function in starburst galaxies
is one of the fundamental goals in studies of starforming regions.
Due to their large luminosity, massive stars are directly visible 
in the integrated spectrum of young starbursts and thereby provide
a unique opportunity to study their stellar content.
In a subset of emission line galaxies -- often referred to as
Wolf--Rayet galaxies (cf.~Conti \markcite{c91} 1991) -- 
{\em broad stellar emission lines} in the optical 
(most prominently \heii\ $\lambda$ 4686)
testify to an important population of Wolf--Rayet (WR) stars, which are 
the descendents of the most massive O stars (see e.g.~Maeder \& Conti
\markcite{mc94} 1994).
The {\em nebular emission lines}, on the other hand, are due to the exciting
stars present in the starburst including the less massive OB stars.
Observations of WR features and nebular lines thus contain information about
stars from different mass ranges. WR galaxies therefore in particular offer a 
unique opportunity to probe the upper part of the IMF in young starbursts.

\subsection{Difficulties with the non-evolving cluster approach}
\label{s_nonevolving}
Since the first discoveries of WR stars in emission line galaxies 
by Allen, Wright, \& Gross \markcite{a76} (1976) and Kunth \& Sargent 
\markcite{ks81} (1981), the recent work of Vacca \& Conti 
\markcite{vc92}(1992, hereafter VC92)
provides the most detailed quantitative study of the massive star 
population in WR galaxies. Their technique, developed in more detail
by Vacca \markcite{v94} (1994, hereafter V94), allows to determine 
the relative number of WR and O stars. 
Comparisons with stellar evolution models show that the
large WR/O ratios can only be explained by {\em bursts of star formation} 
occurring over short timescales compared to the lifetime of massive
stars (Arnault, Kunth, \& Schild \markcite{a89} 1989, 
\markcite{vc92} VC92, Meynet \markcite{m95} 1995).

At first sight the comparison of the WR/O ratios of VC92 \markcite{vc92}
and Contini, Davoust, 
\& Consid\`ere \markcite{cdc95} (1995, hereafter CDC95) 
with the evolutionary calculations of Meynet \markcite{m95} (1995), 
however, seems to show that a {\em flat IMF slope} 
($\alpha \sim$ 1--2, compared to the Salpeter value of $\alpha=2.35$) is 
required to explain the observations (Meynet \markcite{m95} 1995, 
\markcite{cdc95} CDC95)\footnote{It should be noted that 
in Fig.~3 of Maeder \& Conti \markcite{mc94} (1994) the slope of the 
IMF has inadvertently been mislabeled, due to a confusion in 
the notation. 
Their values of $\Gamma=-1$, and $-2$ should correctly refer to 
$\alpha=1$ and 2 in the above notation, where the Salpeter slope is
given by $\alpha=2.35$. See also the original paper by Meynet 
\markcite{m95} (1995) who uses $\alpha=x$.}.
This result is, however, in disagreement with several other 
methods, such as direct stellar counts in clusters 
($\alpha \sim$ 2 -- 3.1; cf.~\markcite{mc94}Maeder \& Conti) 
and population synthesis techniques (Olofsson \markcite{ol89} 1989, 
Mas-Hesse \& Kunth \markcite{mh91} 1991, Kr\"uger \etal \markcite{k92} 
1992, Vacca \etal \markcite{v95} 1995),
which show that observations of similar (and partly even the same)
objects are compatible with a Salpeter IMF over a large range of 
metallicities.

It can, indeed, easily be traced back why applying the method of 
VC92 \markcite{vc92} and V94 \markcite{v94} 
can lead to erroneously large WNL/O ratios, which would imply a 
flat IMF.
From the ratio of the luminosity in the WR feature to the 
\hbeta\ luminosity, they derive 
$N_{\rm WNL}/N^\prime_{\rm O7V}$, the number of WNL stars
over the number of so-called ``equivalent O7V stars'', which are 
required to produce the Lyman continuum flux measured from the
\hbeta\ line.
They then deduce the ratio of WNL over OV stars from
$N_{\rm WNL}/N_{\rm OV} = \eta_0 \,N_{\rm WNL}/N^\prime_{\rm O7V}$,
where $\eta_0$ is the IMF averaged ionizing Lyman continuum 
luminosity of a population normalized to the output of one 
``equivalent'' O7V star (see V94).

The fundamental assumption entering the deri\-va\-tion of
$N_{\rm WNL}/N_{\rm OV}$, or equivalently $\eta_0$, is that the 
Lyman continuum luminosity is produced by an 
{\em unevolved Zero-Age Main Sequence} (ZAMS) population.
Obviously this assumption is not compatible with the presence of
WR stars --- the descendents of O stars --- which clearly indicate 
ages of typically at least 3 Myr. 
$N_{\rm WNL}/N_{\rm OV}$ therefore denotes the number ratio of WNL stars
with respect to a purely theoretical ZAMS OV star population, which: 
{\em 1)} has no observational correspondence, and 
{\em 2)} cannot directly be compared to WR/O number ratios derived 
from evolutionary models.

\begin{figure}[tp]
\centerline{\psfig{figure=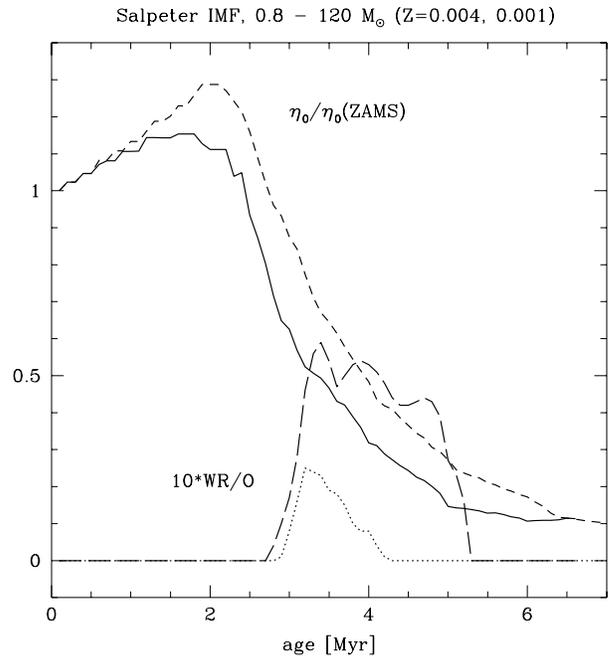,height=8.8cm}}
\caption{Time evolution of $\eta_0$ (solid and short-dashed line)
in an instantaneous burst for Z=0.004 and 0.001 respectively. 
A Salpeter IMF from 0.8 to 120 \msun\ is assumed. 
To indicate the time occurrence of the WR phase
the theoretical WR/O number ratio (long-dashed, dotted for 
Z=0.004 and 0.001) is also plotted here.
At this time the decrease of $\eta_0$ with respect to its ZAMS 
value implies an {\em overestimate} of the WR/O ratio by typically
a factor of $\sim$ 1.1 -- 7.4 for the metallicities shown here, 
depending on the exact age of the population}
\label{fig_1_eta0_change}
\end{figure}

In an instantaneous burst,
due to the evolution of the O star population and the subsequent
disappearance of the most massive stars, the number of 
Lyman continuum photons, $Q_0$, strongly decreases after 2--3 
Myr. This is illustrated in Fig.~\ref{fig_1_eta0_change}, where
we plot the time evolution of 
\begin{equation}
	\eta_0\left(t\right)=
	  \frac{\int_{M_{\rm low}}^{M_{\rm up}} \Phi\left(M\right) 
		Q_0\left(M,t\right) dM}
	  {Q_0^{\rm O7V} \int_{M_{\rm O}}^{M_{\rm up}} \Phi\left(M\right) dM},
\end{equation}
with respect to its ZAMS value $\eta_0(t=0)$ in an instantaneous burst.
Following V94, $\Phi$ is the IMF with the mass limits $M_{\rm low}$ and 
$M_{\rm up}$, $Q_0\left(M,t\right)$ is the ionizing luminosity as a 
function of initial mass and time, and $\log Q_0^{\rm O7V}=49.05$. 
The integral in the denominator counts the number of O stars
defined as having $\teff \ge$ 33000 K.

The strong decrease of $\eta_0$ after $\sim$ 2 Myr implies that the
required {\em number of O stars will be systematically underestimated}
(and hence the WR/O ratio overestimated)
if one does not account for the evolution of the population and
uses the ZAMS values of a ``non-evolving cluster''. 
For younger ages the approach of V94 should, however, be valid.
At the low metallicities typical of the objects from \markcite{vc92}
VC92, the decrease of $\eta_0$ in the WR rich phases (see 
Fig.~\ref{fig_1_eta0_change}) implies 
that the quantity $N_{\rm WNL}/N_{\rm OV}$ overestimates
the real WNL/O ratio by a factor of 1.1 to 7.4,
depending on the exact age of the burst.
%
As can be seen from Meynet \markcite{m95} (1995, Fig.~6), 
such a decrease of the WNL/O ratio allows an important steepening of 
the IMF slope, which may bring the values closer to a Salpeter-like
IMF. 

The possible difficulty of determining the number of O stars from 
$\eta_0$ for ``evolved'' populations was already mentioned by V94. 
One could in principle account (to first order) for the 
evolution by simply lowering the upper mass limit to the value of 
the present-day mass function. 
Such a correction has, however, not been applied by
\markcite{vc92} VC92 and \markcite{cdc95} CDC95.
To obtain a better quantitative handle on the massive star
populations in young starburst galaxies the ideal way is
to synthesize directly the relevant observable quantities
consistently with stellar evolution models. Such an approach
is presented in the following Section.

\begin{figure}
\centerline{\psfig{figure=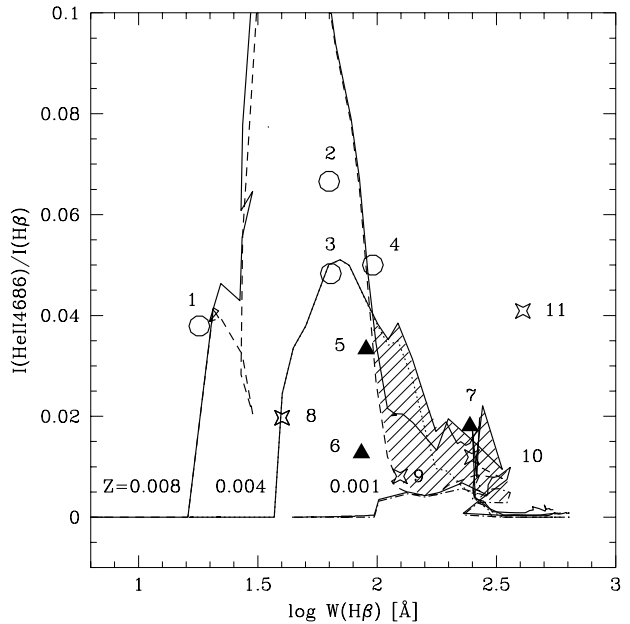,height=8.8cm}}
\caption{Predicted He~{\sc ii}/H$\beta$ intensity as a function
of the H$\beta$ equivalent width in instantaneous bursts at
$Z=$0.008, 0.004, and 0.001.
The solid lines include both the nebular He~{\sc ii} and the broad 
WR emission. The dashed, dotted and dash-dotted lines show the
pure WR emission. The shaded domain illustrates the region, 
where a {\em strong nebular} contribution is predicted.
Observations from VC92 \protect\markcite{vc92} 
and CDC95 \protect\markcite{cdc95} are identified as follows.
1: Mrk 1094, 2: Mrk 712, 3+4: NGC 3125 A+B, 5: Tol 35,
6: NGC 1741 B, 7: II Zw 40, 8: He 2-10 A, 9: Mrk 1236 A, 
10: Pox 4, 11: Pox 139}
\label{fig_2}
\end{figure}

\section{Evolutionary synthesis models for WR galaxies}
\label{s_evol}
We have built population synthesis models using the latest
Geneva stellar evolution tracks (see Meynet \markcite{papv} 
\etal 1994 and references therein) and atmosphere models 
from Kurucz \markcite{k91} (1991) for stars with initial masses 
$M_{\rm ini} < 25 \msun$. 
To obtain a reliable description of the ionizing fluxes
of the hot star population we include the recent spectra 
from the ``combined stellar structure and atmosphere models'' 
({\em CoStar}) of Schaerer \markcite{costar} 
\etal (1996ab) and Schaerer \markcite{crete} (1996) 
for more massive main sequence stars.
WR stars are described by the theoretical spectra of 
Schmutz, Leitherer, \& Gruenwald \markcite{slg92} (1992).
Both the {\em CoStar} and WR atmosphere models account for 
\nlte\ effects and stellar winds, which has a strong influence
on the ionizing flux especially in the \heii\ continuum
(see Schaerer \markcite{costar} et al.).

To study the massive star content in WR galaxies we synthesize 
the following observational quantities:
{\em 1)} \hbeta\ luminosity and \hbeta\ equivalent width
accounting for  stellar and nebular continuum emission,
{\em 2)} nebular \heii\ $\lambda$ 4686 emission, and
{\em 3)} broad WR \heii\ $\lambda$ 4686
	emission feature.
The nebular continuum and recombination lines were derived 
assuming 
$N_e=$ 100 ${\rm cm^{-3}}$, $T_e=$ 10000 K, and solar H/He abundances
(see e.g.~Osterbrock \markcite{o89} 1989).
For the WR feature we synthesize the broad \heii\ $\lambda$ 4686 line
in order to compare our predictions with the high resolution
observations of \markcite{vc92} VC92, which do not include broad 
\ion{N}{3} and  \ion{C}{3} blends at $\lambda \sim$ 4650\footnote{At 
lower resolution these features, often referred to as the ``WR bump'', 
are hardly separable. Emission in the WR bump is larger than in the 
4686 feature alone, and the behaviour of its time evolution differs both 
qualitatively and quantitatively from the pure \heii\ 4686 
feature modeled in this work (cf.~Meynet \markcite{m95} 1995).}.
Consistently with their analysis we adopt an average luminosity of 
$1.7 \, 10^{36} \, {\rm erg \, s^{-1}}$ for WN stars. WC/WO subtypes
have a negligible flux in the \heii\ $\lambda$ 4686 line.

Figure \ref{fig_2} shows the predicted evolution of the 
(WR+nebular) \heii$/ {\rm H}\beta$ intensity as a function of the 
\hbeta\ equivalent width for instantaneous bursts assuming a Salpeter 
IMF ranging from 0.8 to 120 \msun. The results for three metallicities 
$Z$ are plotted. Correspondingly, the observations by VC92 \markcite{vc92}
and CDC95 \markcite{cdc95} are grouped into three metallicity ``bins'':
$0.010 > Z \ge 0.005$ (open circles), 
$0.005 > Z \ge 0.003$ (filled triangles), and
$0.003 > Z \ge 0.0025$ (stars). 
The time evolution proceeds towards low $W(\hbeta)$. 
At $\log W(\hbeta) \sim 2.5$, the increase of \heii$/$\hbeta\
corresponds to the onset of the WR rich phase of the staburst.
Note the strong decrease of the maximum \heii$/$\hbeta\ emission 
with decreasing $Z$ due to strong dependence of the WR population 
on metallicity (cf.~Maeder \& Meynet \markcite{mm94} 1994).
In this diagram adopting a flatter IMF would primarily 
translate to an increase of the \heii$/$\hbeta\ emission at a given
$W(\hbeta)$ by increasing the WR emission. 

Figure \ref{fig_2} clearly shows that, except for object 11,
all observations lie within the range covered by the model
predictions. For each metallicity group the agreement is good.
Most importantly, the observations
for objects 1--5 (see figure caption) which, according to Meynet 
\markcite{m95} (1995) and \markcite{cdc95} CDC95, would require a 
very flat IMF ($\alpha \sim 1$)
can thus very well be explained with a ``standard'' Salpeter IMF.
Such important deviations from the non-evolving cluster approach
are indeed expected, since according to their $W(\hbeta)$ these bursts 
have ages $t >$ 4 -- 4.5 Myr (see Section \ref{s_nonevolving}).
Our results confirm the similarly detailed approach of
Kr\"uger \etal \markcite{k92} (1992). They are also in agreement
with unpublished observations of other WR galaxies (Vacca, private 
communication).

Only two objects (8 and 11) seem to disagree with the Salpeter 
IMF.
He 2-10 (object 8), which has a metallicity of $Z \sim 0.003$, 
might indeed require a somewhat steeper IMF. 
Pox 139 (object 11) appears as an extreme object in the VC92
\markcite{vc92} sample beacause of its large \heii$/$\hbeta\ 
ratio at the young age ($t \sim$ 2.5 Myr) implied the by high 
value of $W(\hbeta)$.
It must, however, be noted that $W(\hbeta)$ is considerably 
uncertain due to the very low continuum flux measured in the 
spectrum of VC92 \markcite{vc92} (Vacca, private communication).
The Pox 139 observations of Kunth \& Sargent \markcite{ks83} 
(1983) show a \heii$/$\hbeta\ value lower by a factor of 1.8.
Better quality observations would be very useful for this galaxy. 

\subsection{Nebular He~{\sc ii} emission in young starbursts}
The presence of nebular \heii\ emission in giant \hii\ regions,
requiring a very hard exciting spectrum, appears puzzling 
(see e.g.~Garnett \etal \markcite{g91} 1991).
The use of the latest {\em CoStar} model fluxes for O stars 
and appropriate atmosphere models for WR stars (see above)
allows us to readdress this question here.

\begin{figure}[htb]
\centerline{\psfig{figure=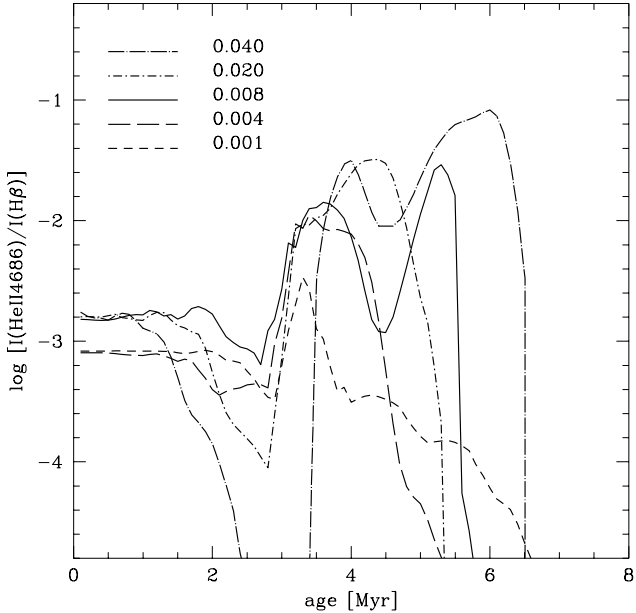,height=8.8cm}}
\caption{Predicted {\em nebular} emission line ratio of 
He~{\sc ii}/H$\beta$
as a function of burst age for different metallicities ($Z=$0.001 
to 0.04) assuming a Salpeter IMF from 0.8 to 120 \msun.
}
\label{fig_4}
\end{figure}

Figure \ref{fig_4} shows the predicted nebular \heii$/$\hbeta\ 
ratio in instantaneous bursts at different metallicities.
First we note that for young bursts we predict typical values
of $I($\heii$)/I(\hbeta)$ between $5. \, 10^{-4}$ and 
$2. \, 10^{-3}$.
Due to the strong \heii\ continuum flux obtained from the 
O star models, which account for \nlte\ effects, line 
blanketing, and stellar winds (cf.~Schaerer \etal 
\markcite{costar} 1996b) 
these values are $\sim$ 4 orders of magnitudes larger than 
the values which would be derived from other current 
population synthesis models. 

After $\sim$ 3 Myr the \heii$/$\hbeta\ ratio increases due to the
appearance of WR stars. 
The major nebular emission is due to stars in the WC phase.
Since these stars have a negligible broad stellar emission in the
4686 line, we thus expect an {\em important nebular contribution 
to the total 4686 emission in the early WC-rich phase}.
The domain corresponding to this phase is shown by the
hatched area in Fig.~\ref{fig_2}.
Typical values of $I($\heii$) / I(\hbeta) \sim$
0.01 -- 0.025 are attained.
At later times (characterized by $W(\hbeta) <$ 100 $\rm \AA$
in Fig.~\ref{fig_2}), the broad stellar component from the then
important WNL population will dominate the 4686 emission.
The strongest relative nebular contribution is predicted between 
solar metallicity and $\sim 1/5 \,Z_\odot$,
where in the early WR phase the nebular flux can make up to 
$\sim$ 75 \% of the emission in the \heii\ 4686 line. 
At $Z > 0.02$ and $Z < 0.004$ the broad WR feature always
dominates the 4686 emission\footnote{At high $Z$ the reduction 
of the absolute \hbeta\ emission overwhelms the increase 
of the nebular \heii$/$\hbeta. At low $Z$, due to the
low mass loss, the WC population becomes negligible.
This explains the dominance of the  the broad WR emission
for these metallicities.}.
 
First comparisons with observations show the following:
{\em 1)}
Two objects in Fig.~\ref{fig_2} lie in the domain where the 
nebular contribution to the total WR+nebular emission is expected 
to be high.
Our prediction is confirmed by inspection of the spectrum of Pox 4 
from VC92 \markcite{vc92} (their Fig.~6), which indeed shows an 
important nebular contribution. 
We expect a similar nebular contribution in II Zw 40. 
We note that these objects also show the largest electron temperature 
in the VC92 \markcite{vc92} sample, which is an additional 
confirmation of their high excitation.
For a more firm conclusion about the excitation conditions
in Mrk 712 higher resolution data should be obtained.
{\em 2)} Of the 12 objects from Campbell, Terlevich, \& Melnick 
\markcite{c86} (1986) showing nebular \heii\ emission four 
(T 1304-353, T 1304-386, C 1543+091, F 30) 
can well be explained by our models.
We suspect that many of the remaining objects might well turn
out to show broad \heii\ emission if observed with a sufficient
signal/noise (see also Masegosa, Moles, del Olmo \markcite{m91} 
1991). Their three additional objects with broad WR features
are also compatible with our models.

\section{Summary and outlook}
We have shown that methods, which do not account for the evolution
of the bursting population will {\em systematically underestimate}
the number of O stars derived from integrated spectra for burst
ages $t > 2 - 3$ Myr. 
As a consequence, present WR/O number ratios in WR galaxies 
\markcite{vc92} (VC92, \markcite{cdc95} CDC95) are thus overestimated.
Therefore IMF determinations based on such number ratios will 
preferentially lead to flat IMF slopes ($\alpha \sim$ 1--2;
Meynet \markcite{m95} 1995, \markcite{cdc95} CDC95).
If one instead properly treats the evolution,
the observed massive star population in WR galaxies is compatible 
with resulting from an instantaneous burst following a Salpeter IMF,
in agreement with previous studies using different methods.
This conclusion may, however, need to be revised if other systematic
uncertainties (photon leakage, dust, slit positioning etc.; see
e.g.~Conti \markcite{c93} 1993) turned out to be important.
Obviously it must also be noted that the observed properties
are probably not uniquely described by the case of an 
instantaneous burst discussed here, and that several uncertainties
remain (e.g.~ionizing fluxes of WR stars).

Using recent stellar evolution models and new spherically 
expanding \nlte\ atmosphere models for O and WR stars, which in 
particular yield reliable predictions for the ionizing flux in 
the \heii\ continuum, we have proposed a natural explanation for 
the observation of nebular \heii\ in extragalactic \hii\ regions 
and alike objects:
High excitation objects should be intimately linked with an important 
population of WC stars.
It is indeed feasible to test this prediction
by detecting the broad \ion{C}{4} $\lambda$ 5808 emission line 
characteristic of WC stars (see also Meynet \markcite{m95} 1995).
In relatively short phases ($\sim$ 1 Myr) the 
expected emission (using the WC fluxes from Smith \markcite{s91}
1991) reaches up to $I($\ion{C}{4}$)/ I(\hbeta) \sim$ 0.02 -- 0.2 
for metallicities between 1/5 $Z_\odot$ and solar.
Similarly, a careful analysis of the \ion{C}{3} $\lambda$ 4650 
blend (cf.~Kr\"uger \etal \markcite{k92} 1992) and maybe even 
the \ion{C}{3} $\lambda$ 5696 line could also be used to measure 
the WC population.

Although little is presently known about WC stars in starbursts
the observations of Gonz\'{a}lez-Delga\-do et al. (19\-94) 
\markcite{g94} showing strong nebular \heii\ emission associated 
with WC stars support our picture.
Future studies should draw their attention on obtaining
high quality observations (and providing detection/non-detection
limits !) encompassing both the ``traditional'' WR bump
($\lambda$ 4650 -- 4686) with a sufficient spectral resolution,
and the C emission lines. 
Predicted starburst spectra allowing detailed probes of WN and WC 
populations will be presented elsewhere (Schaerer \& Vacca 1996, in 
preparation).
A systematic study of a large sample of young starbursts, including
higher metallicity objects, will provide crucial tests
for stellar evolution and the understanding of star formation 
processes in these environments.

\acknowledgments
Its a pleasure to thank Thierry Contini, Geor\-ges Meynet,
Daniel Kunth, Rosa Gonz\'alez-\-Del\-gado, Tim Heckman, Claus Leitherer,
and especially Bill Vacca for stimulating discussions. 
I would also like to thank the people at the Observatory in Toulouse
who made my stays possible.
This research in part supported by the Swiss National
Foundation of Scientific Research.
Partial support from the Directors Discretionary 
Research Fund of the STScI is also acknowledged.

\end{document}